# Causal Structures of Dynamic Black Holes[a]

Beth A. Brown[#*] and James Lindesay[*]

[#]*NASA Goddard Space Flight Center, Greenbelt, Maryland*
[*]*Computational Physics Lab, Howard University, Washington, D.C.*

**Abstract.** Dynamic space-times, especially those manifesting horizons, provide useful laboratories for examining how macroscopic quantum behaviors consistently co-generate gravitational phenomena. For this reason, the behaviors and large-scale causal structures of spatially coherent dynamic black holes will be explored in this presentation. Geodesic motions on an evaporating black hole will also be presented. Research recently completed with Beth Brown, including her final Penrose diagram for an accreting black hole, will be presented.

## DYNAMIC SPACE-TIMES

Dynamic black holes have been found to have qualitatively different behaviors from their static counterparts. These differences result from the choice of the temporal coordinate used to describe the dynamics, as well as the form of the metric. It is particularly convenient to develop spatially coherent geometries upon which quantum systems have straightforward behaviors and interpretations. Quantum measurability constraints can be most directly satisfied in dynamic space-time backgrounds.

As an example, consider a rotationally dynamic space-time, represented by the Kerr metric. For brevity, it is convenient to define dimensional length scales identified with the mass and angular momentum of the geometry, given by

$$R_M \equiv \frac{2G_N M}{c^2} \quad , \quad R_J \equiv \frac{J}{Mc}. \tag{1}$$

The form $R_M$ will be referred to as the radial mass scale for the system, and for a static rotationally symmetric system, it is identified with the Schwarzschild radius. The Kerr metric takes the form

$$ds^2 = -\left(1 - \frac{R_M r}{r^2 + R_J^2 \cos^2 \theta}\right)c^2 dt^2 + \frac{r^2 + R_J^2 \cos^2 \theta}{R_J^2 + r(r - R_M)} dr^2 +$$

$$\left(r^2 + R_J^2 \cos^2 \theta\right)d\theta^2 - 2\frac{R_M R_J \sin^2 \theta}{r^2 + R_J^2 \cos^2 \theta} c\, dt\, r\, d\varphi + \quad (2)$$

$$\left(r^2 + R_J^2 + \frac{R_M R_J^2 r \sin^2 \theta}{r^2 + R_J^2 \cos^2 \theta}\right)\sin^2 \theta\, d\varphi^2.$$

The metric is seen to have a non-orthogonal temporal/angular term associated with the rotational dynamics of the geometry. This term contributes to frame dragging behaviors in the space-time. The metric is generally valid in the exterior region even if there is no black hole present.

One might construct a similar form to describe a radially dynamic spherically symmetric space-time. For a static geometry, a form that has been put forth by the *river model* of black holes[1] gives some insight into how this can be done. If a dependency of the radial mass scale on this temporal coordinate is incorporated, the dynamic metric has the form

$$ds^2 = -\left(1 - \frac{R_M(ct)}{r}\right)c^2 dt^2 + 2\sqrt{\frac{R_M(ct)}{r}}\, c\, dt\, dr + dr^2 + r^2 d\theta^2 + r^2 \sin^2 \theta\, d\varphi^2. \quad (3)$$

This geometry introduces no physical singularity near the horizon (if the vacuum solution continues to be valid down to that region). The mixed temporal-radial term in this radially dynamic metric is analogous to the mixed temporal-angular term in the rotationally dynamic Kerr metric. The metric is seen to have asymptotic correspondence to that of flat Minkowski space-time, indicating that the temporal coordinate is that of a distant observer, analogous to the temporal coordinate in a Schwarzschild space-time that represents the time of a distant Schwarzschild observer. However, although both asymptotic spaces are Minkowski, it should be noted that they are not the same Minkowski spaces[2].

## PENROSE DIAGRAMS

Penrose diagrams are convenient tools for exploring the large-scale causal structure of a particular space-time. These diagrams are space-time diagrams that preserve the slope of light-like trajectories to be ±unity (as in Minkowski space-time), while mapping the entire infinite domains of the geometry onto a finite region. Because of the ubiquitous behaviors of light-like trajectories, causal relationships on such diagrams can be ascertained in a direct manner. For spherically symmetric space-times, there is no angular dependency on relationships, so that angular coordinates can be suppressed. For such Penrose diagrams, any point on the diagram represents a spherical surface with the given radial coordinate at the given time. Example Penrose diagrams will be constructed in what follows.

# Minkowski Space-Time

The metric of a flat Minkowski space-time takes the following form in spherical polar coordinates:

$$ds^2 = -c^2 dt^2 + dr^2 + r^2 d\theta^2 + r^2 \sin^2\theta \, d\varphi^2. \qquad (4)$$

One is probably familiar with the standard space-time plot of stationary observers and light beams given in Figure 1.

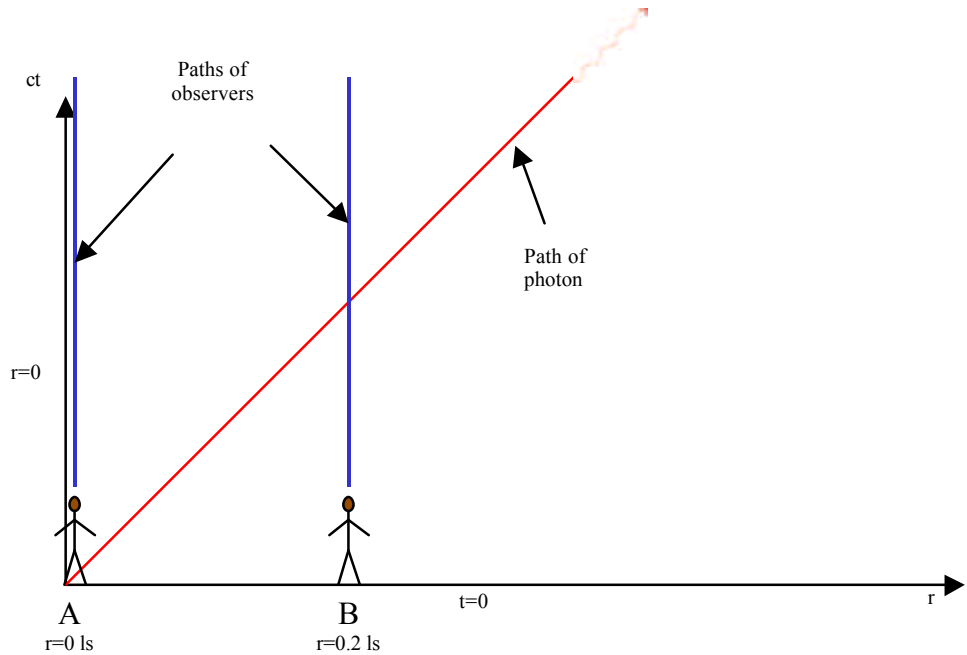

**FIGURE 1.** Standard Minkowski space-time diagram

The stationary observers A and B are separated just beyond the distance of the orbits of geo-synchronous communication satellites. A light beam emitted by observer A is received by observer B 0.2 seconds later. No events above this light-like trajectory can have causal influence upon an event below this path. The diagram is convenient and intuitive, however, the observer at 1.0 light-second is beyond the region of display of the diagram.

In order to display the entire Minkowski space-time onto a finite region, one must choose a functional form that maps infinity onto a finite number. For the diagrams displayed in this presentation, the functional form will always be chosen to be hyperbolic tangents, which map $\pm\infty$ onto $\pm 1$. The Penrose horizontal and vertical coordinates are chosen to be of the form

$$Y_h = \frac{-Tanh(ct-r)+Tanh(ct+r)}{2}$$
$$Y_v = \frac{Tanh(ct-r)+Tanh(ct+r)}{2}.$$
(5)

For Minkowski space-time, the Penrose diagram is then given in Figure 2:

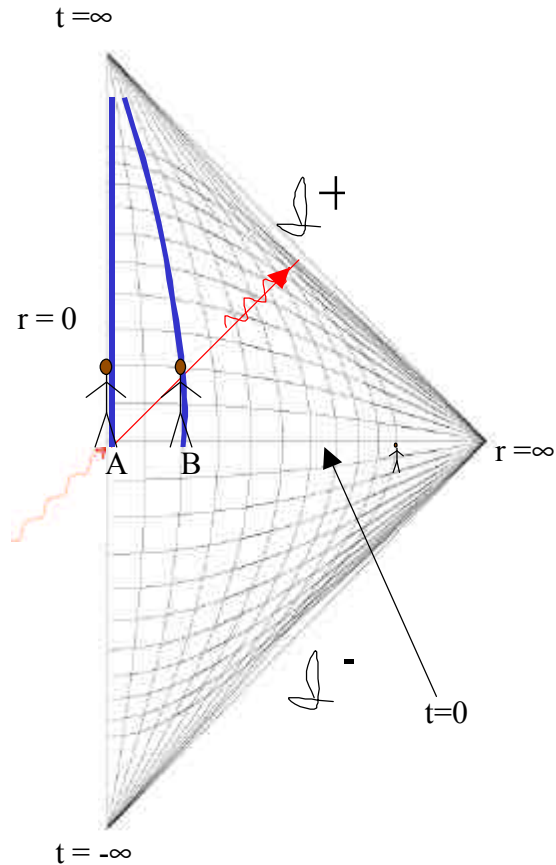

**FIGURE 2.** Penrose diagram for Minkowski space-time.

Any given point *(r,ct)* on the Penrose diagram represents a sphere with radius *r* at time *t*. Both observers A and B, as well as the observer located at 1.0 ls, are displayed on the diagram. In fact, the whole of the space-time is displayed, although the scale of the equally sized observers is location dependent. One important feature of the diagram is that the light-like communication from A to B has a preserved slope, defining potential causal relationships on the global space-time. Also, past infinity is represented as an outgoing light-light surface *skri⁻*, while future infinity is represented as an ingoing light-like surface *skri⁺*. These characteristics of the *Asymptopia* of Minkowski space-time will be preserved in the dynamic geometries that follow.

# Excreting Black Hole

Next, the global causal structure of an evaporating black hole satisfying a metric form given in Eq. (3) will be examined. Conformal coordinates have been developed for the case of steady evaporation[2]. Various useful properties of the resulting Penrose diagram can then be directly examined.

## *Dynamic Coordinate Grid*

The Penrose diagram of the radially dynamic evaporating black hole is given in Figure 3.

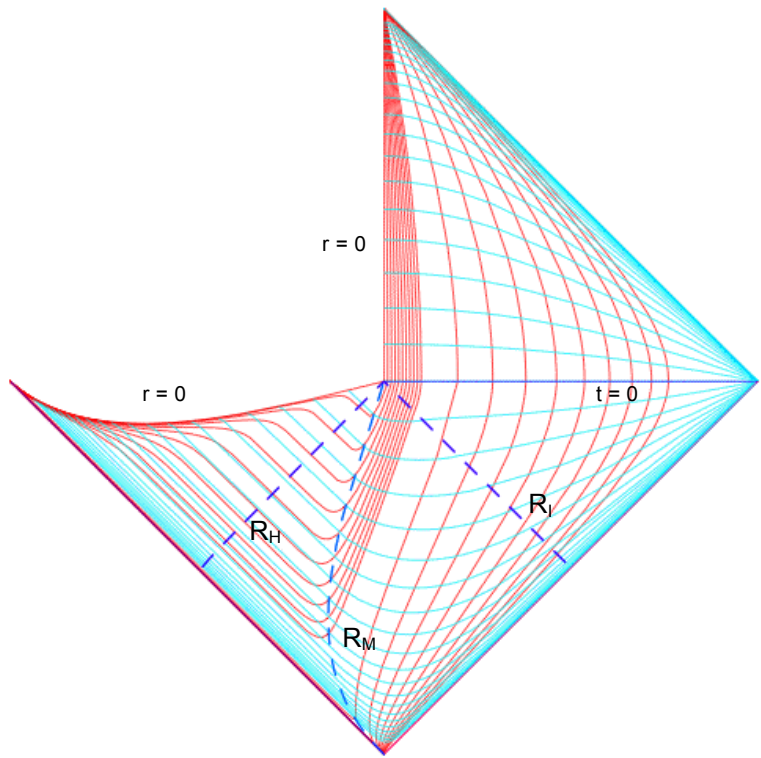

**FIGURE 3.** Penrose diagram of a spatially coherent black hole that evaporates at a fixed rate.

Any given point *(r,ct)* on the Penrose diagram represents a sphere with area *4**p** r²* at time *t*. The space-like singularity at r=0 evaporates away at t=0 to become the low-curvature time-like origin which is non-singular. The low-curvature upper right-hand region of the diagram has the same global structure as Minkowski space-time. The fixed radial coordinate curves that run vertically (time-like) in the upper right portion of the diagram are initially graded in tenths, then in units of the given scale. The fixed temporal coordinate curves that run horizontally (space-like) in the diagram are graded in units of the given scale.

The dynamic horizon is the out-going light-like surface labeled $R_H$. The radial mass scale is labeled $R_M$, and in contrast to the static Schwarzschild radius that crosses no temporal curves, the dynamic radial mass scale and horizon cross temporal and radial coordinate curves. Outgoing photons are temporarily stationary in the radial coordinate at the radial mass scale, and all light-like or time-like trajectories within this surface must necessarily have decreasing radial coordinate. The in-horizon $R_I$ is the last ingoing light-like surface that can communicate with the singularity.

## *Causal Relationships*

The global causal structure of the evaporating black hole can be examined using 5 example events with coordinates A, B, C, D, and E demonstrated in Figure 4.

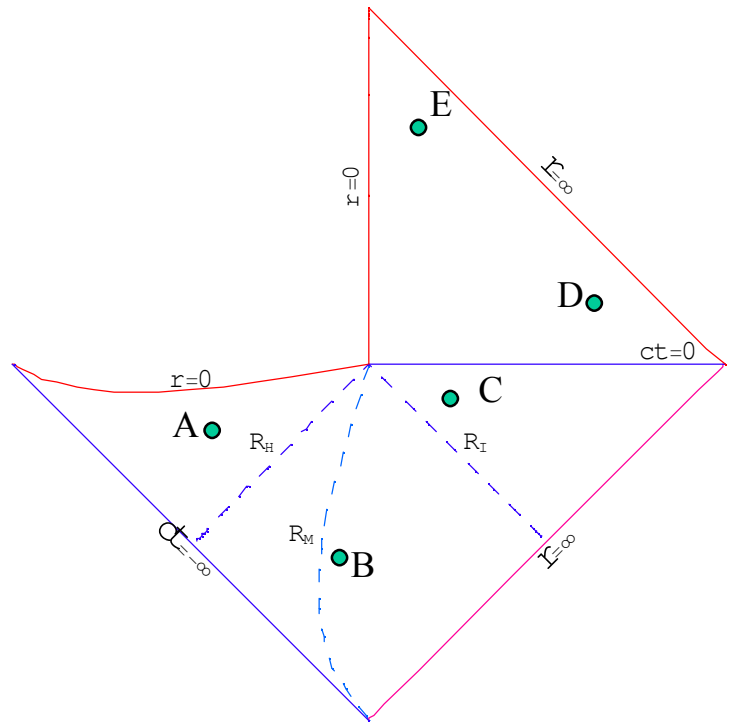

**FIGURE 4.** Example events on the Penrose diagram of an evaporating black hole.

Event A occurs within the horizon, event B just outside the radial mass scale, event C outside of the in-horizon, and events D and E occur after the singularity has evaporated away. The various causal relationships are demonstrated in the diagram in Figure 5.

|   | A | B | C | D | E |
|---|---|---|---|---|---|
| A |   | In Light Effect | Space-like | Space-like | None |
| B | In Light Cause |   | Time-like Cause | Out Light Cause | Time-like Cause |
| C | Space-like | Time-like Effect |   | Space-like | Time-like Cause |
| D | Space-like | Out Light Effect | Space-like |   | In Light Cause |
| E | None | Time-like Effect | Time-like Effect | In Light Effect |   |

**FIGURE 5.** Causal relationships of events on evaporating black hole geometry.

The table indicates that, for example, event B can be the cause of an ingoing light-like communication that affects A. Space-like relationships, like those between A and C, can represent space-like coherent behaviors, like those between quantum entangled events. Of particular interest is the relationship between A and E, which is neither time-like, light-like, or space-like. This means that no primary (single) process or wavefunction can entangle these events.

## Geodesic Motions

Freely gravitating objects in the metric Eq. (3) satisfy the geodesic equations for this geometry. The geodesic equations are given by

$$\frac{du^{ct}}{dl} = -\frac{z^{1/2}}{2r}\left(z^{1/2}u^{ct} + u^r\right)^2,$$

$$\frac{du^r}{dl} = -\frac{\dot{R}_M}{2rz^{1/2}}\left(u^{ct}\right)^2 - \Theta_m \frac{z}{2r}, \qquad (6)$$

$$\frac{dz}{dl} = \frac{\dot{R}_M u^{ct} - z\, u^r}{r},$$

where the parameter $\Theta_m$ is unity for massive particles and vanishes for massless particles. The relationship between components of the four-velocities can be obtained from the metric, and is given by

$$u^r = -\sqrt{\frac{R_M(ct)}{r}}\, u^{ct} \pm \sqrt{(u^{ct})^2 - \Theta_m} \,. \qquad (7)$$

The trajectories for light-like objects have been developed in prior work[3]. Example trajectories for massive freely gravitating particles are demonstrated on the Penrose diagram in Figure 6.

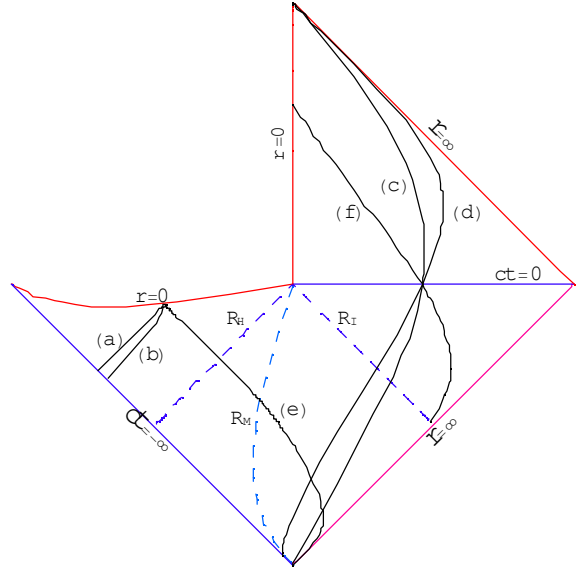

**FIGURE 6.** Geodesic trajectories of outgoing and ingoing massive particles.

Trajectories (a) and (b) represent outgoing particles with differing speeds within the horizon that hit the singularity at a chosen time, while trajectory (e) represents an ingoing particle that falls through the horizon and hits the singularity at that same time. Trajectories (c) and (d) represent outgoing particles that escape to future infinity after passing through a given radial coordinate at time t=0 with differing speeds, while trajectory (f) represents the trajectory of an ingoing particle that originates external to the in-horizon, passes through that same radial coordinate at t=0, and later reaches the center at a later finite time.

The metric (3) admits a special type of geodesic trajectory. Because of the non-orthogonal form of the metric, the relationship between the radial and temporal components of the four-velocities is given in Eq. (7). There are solutions of this equation for massive particles which are neither outgoing nor ingoing. For these geometrically stationary solutions the proper times of the stationary observers are the same at that of the asymptotic observer, and the components of the four-velocities satisfy

$$u^{ct} = 1 \,, \quad u^r = -\sqrt{\frac{R_M(ct)}{r}} \,. \qquad (8)$$

The integral of each of these solutions smoothly connect to radially stationary trajectories after the radial mass scale vanishes. Example classical trajectories of geometrically stationary particles are demonstrated in Figure 7.

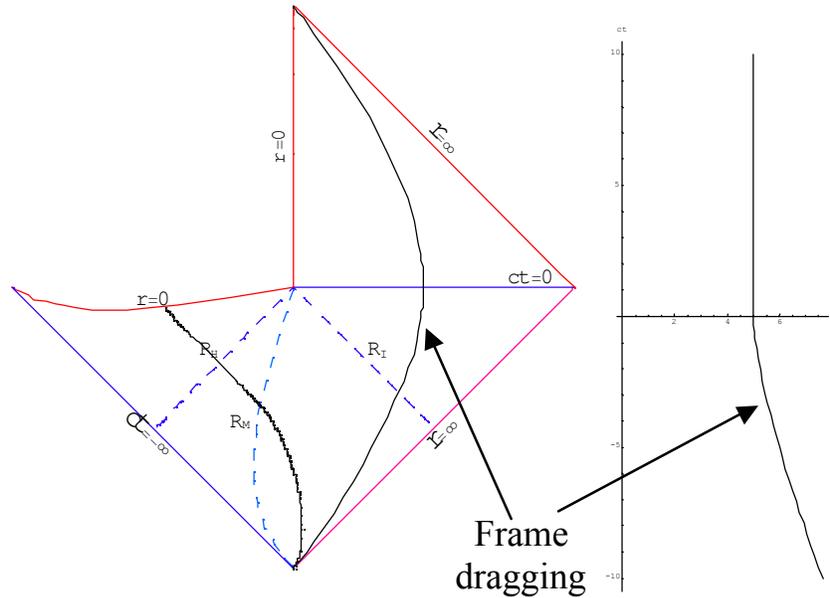

**FIGURE 7.** Penrose and standard space-time diagrams of geometrically stationary trajectories.

The diagram on the left in this figure is a Penrose diagram demonstrating the classical trajectories of a geometrically stationary massive particle that eventually hits the singularity at r=0, and another geometrically stationary massive particle that becomes radially stationary after t=0. It should be noted that the latter stationary trajectory mirrors the flat space form in the upper right even while the black hole is still present in the lower right portion of the diagram. The diagram on the right demonstrates a standard (r,ct) coordinate plot of the latter stationary massive particle, showing that prior to the completion of the evaporation of the black hole, the radial coordinate of the particle is decreasing. However, from Eq. (8), it is clear that the proper time of any geometrically stationary observer is the same as that of an asymptotic observer, despite local curvature.

## Accreting Black Hole

The equations that were utilized to develop the conformal coordinates needed to construct Figure 3 for an evaporating black hole can likewise be used to examine the large scale causal structure of a steadily accreting black hole[4] by choosing a constant positive accreting rate $\dot{R}_M > 0$. The Penrose diagram that results is shown in Figure 8.

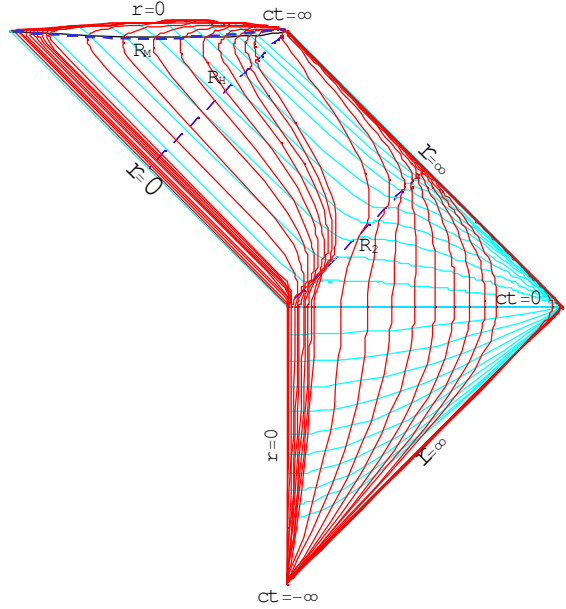

**FIGURE 8.** Penrose diagram for a steadily accreting spatially coherent black hole.

As was the case for the Penrose diagram of the evaporating black hole, the fixed temporal coordinate curves run horizontally (space-like volumes) and are graded in units of the given scale. This is one of the key features of the spatially coherent geometries being investigated. The fixed radial coordinate curves run vertically (time-like) on the right-hand portion of the diagram, and are initially graded in hundredths, tenths, then units, and decades of the given scale.

The time-like center of the flat geometry r=0 is seen to undergo a light-like transition as the accretion begins at t=0 in order to expand the conformal space as needed to form the space-like singularity of the black hole. This demonstrates how the overall geometry enlarges to accommodate the newly forming black hole. The singularity r=0 lies within the radial mass scale $R_M$, which lies within the dynamic horizon $R_H$. All of these features lie within the expanded region of the space-time. There is an outgoing light-like surface labeled $R_2$ along which the coordinates behave in an anomalous manner. This surface (coherently) communicates the initiation of the accretion into a black hole throughout the space-time. The surfaces (r→∞, t→±∞) bounding the diagram on the right represent the Minkowski space-time *asymptopia* of this geometry.

# LIFE CYCLE OF A BLACK HOLE

The previous sections have examined the global causal structure of a black hole that evaporates away from past infinity at a constant rate, and that of a black hole that starts a perpetual accretion at a constant rate. It is of interest to examine the large scale causal structure of a space-time that has a black hole present only for a finite time interval. Each of those geometries involved transitions of the black hole geometries to/from low curvature space-times, matching the metric forms in the volumes at the times of transition. One expects that the Penrose diagram for the complete life cycle of a spatially coherent black hole should be obtained by combining a geometry that begins an accretion with one that ends an evaporation, matching the forms of the metric at the transition from accretion to evaporation[5].

The diagram on the left of Figure 9 represents the original form of the Penrose diagram for the complete life cycle of a black hole that was anticipated by the authors[6,7]. This diagram contains the expected features resulting from a spherically symmetric ingoing light-like shell of finite thickness, eventually falling to a scale within its own Schwarzschild radius, then forming a space-like singularity at r=0.

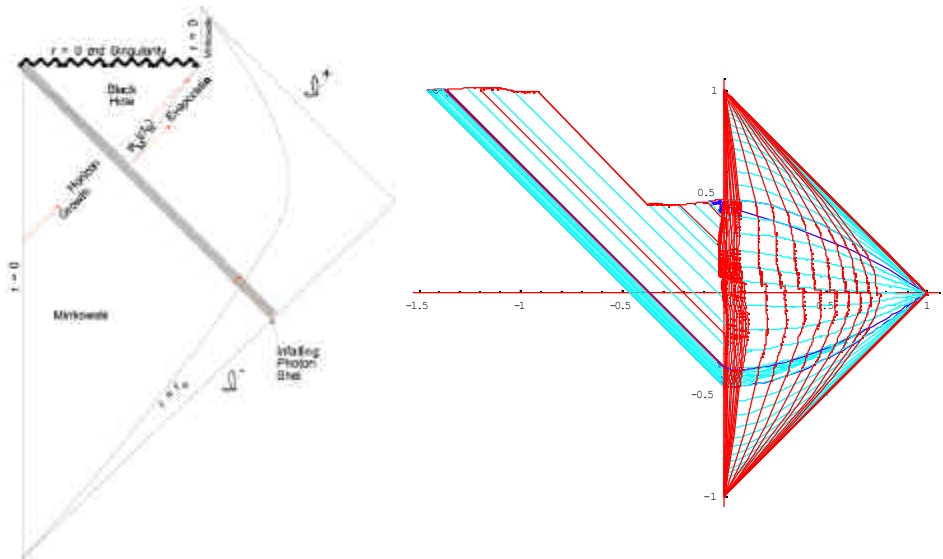

**FIGURE 9.** Penrose diagram of anticipated black hole life-cycle vs. preliminary calculation of spatially coherent black hole.

The anticipated horizon was globally defined by the last outgoing light-like surface that hits the evaporating singularity, while the radial mass scale $R_M$ was anticipated to be a space-like surface internal to the horizon during the accretion of the light-like shell, becoming a time-like surface external to the horizon during evaporation. The Minkowski past infinity surface *skri*$^-$ internal to the ingoing photon shell was anticipated to have a differing scale from the Minkowski future infinity surface *skri*$^+$ after all radiations from evaporation have reached future infinity. Thus, the past and future asymptopias undergo scale changes along the conformal surfaces defined by the ingoing and outgoing causal shells associated with the finite duration of the black hole.

The diagram on the right of Figure 9 demonstrates preliminary calculations of the Penrose diagram of the complete life cycle of a spatially coherent black hole[5]. The times of the initiation of accretion and end of evaporation were chosen symmetrically about t=0 for this example. There are several features of interest on this diagram. First, the right boundaries are those of a Minkowski space-time unaffected by the presence of this black hole of finite duration. Also, the overall geometry is seen to expand to the left of the diagram during the initiation of accretion to accommodate the structures inherent to the black hole through a light-like transition of the time-like center r=0 to the space-like accreting singularity r=0. This singularity undergoes a rapid transition to the space-like evaporating singularity, which evaporates away to once again become a time-like low-curvature center r=0. These features are generically incorporated in the Penrose diagram in Figure 10.

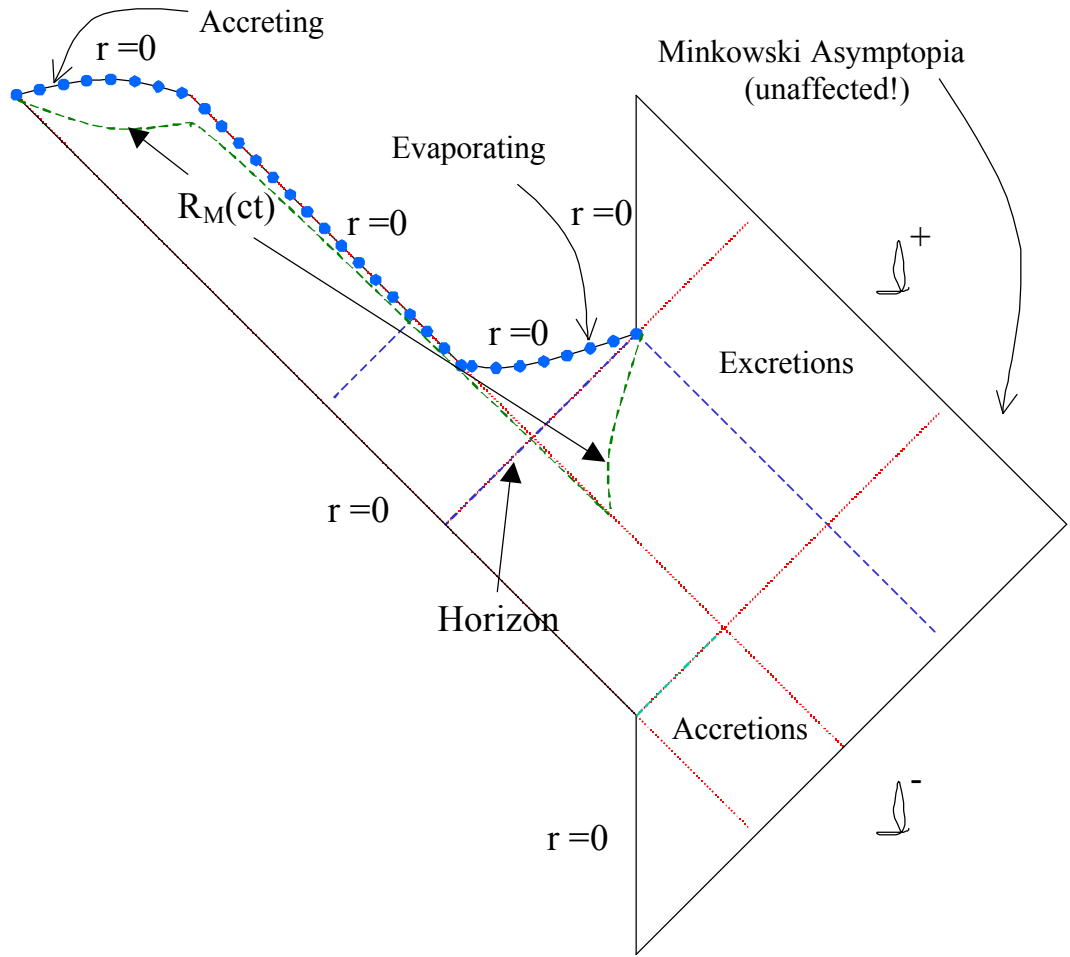

**FIGURE 10.** Expected Penrose diagram for the complete life cycle of a black hole.

The diagram is bounded on the right by the light-like past and future infinities (*skrī⁻* and *skrī⁺*) of the asymptotic Minkowski space-time of this geometry. At the time of the initiation of singular accretion to form the black hole, the center of accretion (which bounds the lower portion of the diagram on the left) undergoes an ingoing light-like transition, expanding the conformal space as needed to accommodate the new black hole. This process should not be confused with any prior accretions within the low-curvature space-time that might have led to "normal', non-singular structures such as stars, white dwarfs, etc. The space-like singularity of the black hole r=0 is represented by the dotted curve that bounds the left-hand portion of the diagram from above. During the period for which accretions exceed excretions, the vertical Penrose coordinate of this singularity has a larger value than it does during the period for which excretions exceed accretions. After the singularity has evaporated away, the center reverts to the time-like, non-singular trajectory represented by the upper vertical line labeled r=0.

The horizon is the (globally constructed) last outgoing light-like surface that reaches the singularity prior to its vanishing into low-curvature space-time. The horizon for the spatially coherent black hole also represents a singular surface for the conformal scale factors during the evaporation stage, but is different from the corresponding singular surface during the accretion stage, which is the outgoing light-like surface just to the left of the horizon that would have terminated at the uppermost future infinity point had accretion continued. The radial mass scale $R_M(ct)$ is expected[6] to remain inside of the horizon during accretion and outside during evaporation. From Eq. (3), outgoing/ingoing null surfaces satisfy

$$\frac{dr_g}{dct} = -\sqrt{\frac{R_M}{r_g}} \pm 1. \qquad (9)$$

Therefore, outgoing light-like trajectories are stationary in the radial coordinate at the radial mass scale. This means that any trajectory within this scale necessarily has decreasing radial coordinate. Generic causal relationships between events in the space-time can be directly ascertained using Figure 10.

## CONCLUSIONS AND DISCUSSION

Penrose diagrams are very convenient constructs for displaying the large-scale causal structure of given space-times. One can conclude that tools for the construction of the Penrose diagrams for dynamic black holes have been successfully developed using non-orthogonal temporal coordinates. Radially dynamic geometries share non-orthogonal temporal-spatial metric forms with rotationally dynamic geometries, such as Kerr space-time. The spatially coherent geometries that have been examined are extremely convenient for exploring the behaviors of quantum systems, because of the dynamic form of the geometries, as well as inherent space-like coherence of quantum systems. The large-scale causal structures of a given space-time provide information on regions of coherence, causal boundaries, and information deficits. The exploration of quantum behaviors on the spatially coherent geometries here examined will be the subject of future presentations.

## ACKNOWLEDGMENTS

This research was supported in part through the NASA Administrators Fellowship Program. The work presented was completed just prior to the untimely passing of one of the co-authors, Dr. Beth A. Brown. The authors gratefully acknowledge the efforts of the session organizer, Chanda Rosalyn Sojourner Prescod-Weinstein, as well as useful discussions with Tehani Finch, Paul Sheldon, and Lenny Susskind.